\documentstyle[preprint,aps,floats]{revtex}
\tightenlines
\input psfig.tex
\begin{document}
\def\ba{\begin{eqnarray}}
\def\ea{\end{eqnarray}}
\def\be{\begin{equation}}
\def\ee{\end{equation}}
\def\({\left(}
\def\){\right)}
\def\[{\left[}
\def\]{\right]}
\def\lagrange {{\cal L}}
\def\del {\nabla}
\def\d {\partial}
\def\Tr{{\rm Tr}}
\def\half{{1\over 2}}
\def\fourth{{1\over 8}}
\def\bibi{\bibitem}
\def\S{{\cal S}}
\def\xx{\mbox{\boldmath $x$}}
\newcommand{\labeq}[1] {\label{eq:#1}}
\newcommand{\eqn}[1] {(\ref{eq:#1})}
\newcommand{\labfig}[1] {\label{fig:#1}}
\newcommand{\fig}[1] {\ref{fig:#1}}
\def\gsim{ \lower .75ex \hbox{$\sim$} \llap{\raise .27ex \hbox{$>$}} }
\def\lsim{ \lower .75ex \hbox{$\sim$} \llap{\raise .27ex \hbox{$<$}} }
\newcommand\bigdot[1] {\stackrel{\mbox{{\huge .}}}{#1}}
\newcommand\bigddot[1] {\stackrel{\mbox{{\huge ..}}}{#1}}
\title{Stability of Flat Space to Singular Instantons
} 
\author{
Neil
Turok\thanks{email:N.G.Turok@damtp.cam.ac.uk}}
\address{
DAMTP, Silver St, Cambridge, CB3 9EW, U.K.}
\date\today 
\maketitle

\begin{abstract}
Hawking and the author have proposed a class of singular,
finite action instantons for defining the initial conditions for
inflation.
Vilenkin has argued they are unacceptable. He exhibited 
an analogous class of 
asymptotically flat instantons which on the face of it
lead to an instability of
Minkowski space. However, all these instantons must be 
defined by introducing a constraint into the path integral,
which is then integrated over. I show that with a careful
definition these instantons do not possess a negative mode.
Infinite flat
space is therefore stable against decay via
singular instantons.

\end{abstract}
\vskip .2in

\section{Introduction}

Hawking and the author recently discovered a new class of 
instantons which could be relevant for inflationary cosmology.
For a generic inflationary potential there exists
a one parameter family 
of finite action solutions to the Euclidean field 
equations, which are the natural deformation of the four sphere solution
to pure gravity with a cosmological constant. They may be
analytically continued to give an open inflating universe.
These instantons possess a singular boundary of zero size, but
it was argued in \cite{ht1,ht2} that 
quantum fluctuations are
well defined in its presence. 
The expectation was expressed that
low energy phenomena would still be uniquely determined in 
the presence of the singularity. This has subsequently
been confirmed  by calculations of the 
perturbation spectra \cite{gratton,hertog}.

Vilenkin has criticised of the use of such
instantons \cite{vil}. He constructed
a related class of
asymptotically flat singular  
instantons which on the face of it lead to the nucleation of
holes in flat space. He 
estimated the rate per unit spacetime volume and
concluded it could be arbitrarily large. As a result, he
advocated excluding such instantons just because they are 
singular. 
I don't think this is a tenable conclusion.
Typical field configurations
contributing to the path integral are 
non-differentiable Brownian random walks,
so to ignore singular configurations is
to ignore essentially everything. 
If the Euclidean action for quantum 
gravity cannot by itself suppress such horrors as the decay of flat space,
it is unlikely that the gravitational path integral can be made sense of
at all. However, a
clue to the correct interpretation of these instantons was 
an observation due to Wu \cite{chao} that even though they 
are solutions of the field equations (away from the
singularity), they are not stationary
points of the action. Their action depends linearly on 
a parameter which governs the strength of the singularity.
Therefore they must 
be treated as constrained instantons \cite{ht2}.

In this Letter I show that a careful treatment of constrained 
singular instantons
removes Vilkenkin's  instability. The simplest argument that
flat space is stable employs conservation of energy. 
The positive energy theorem, 
states that any {\it regular} asymptotically flat
space which has zero ADM energy is flat\cite{sy}. If energy is conserved,
flat space cannot decay. However, Vilenkin's instantons are
singular and have zero ADM energy \cite{Reall}, so this argument 
cannot unfortunately be applied.

We must therefore 
compute the 
the amplitude for flat three space to propagate into itself.
If this possesses an imaginary part we can conclude that
flat space is unstable. As usual we
Wick rotate the path integral to Euclidean time, and 
sum over asymptotically flat Euclidean four geometries.
The only regular solution of the
Einstein equations with these boundary conditions is flat space.
That is the only genuine instanton and  it
posesses no negative mode \cite{gpy}. 

What about fluctuations about 
constrained instantons?  Think of the four geometry
space as a fluctuating rubber sheet. 
If we push on a sheet with a ring of wire, we can deform it. 
The region of the sheet outside the ring
is a nontrivial solution of the 
equations of motion, and the region inside it will be flat.
One can take the radius of the ring to zero, leaving 
a `spike' on the surface of the sheet. I claim that this is the
correct interpretation of Vilenkin's  instantons. Note that
even though
there is nothing to `push' on spacetime, configurations like this
are inevitably produced as quantum fluctuations. We should 
then check for stability around these `spiky' configurations.

It is easy to see that such configurations inevitably
contribute to the
path integral. I am not saying that
they are a significant contribution, nor that they lead
to any sensible approximation scheme for it. 
But they are {\it there}. The point is that one can simply
introduce the identity $1=\int dC \delta ({\cal C}-C)$  into
the functional integral, for some appropriate operator ${\cal C}$.
The simplest example is for massive scalar field theory, for which 
${\cal C}$ could just be the value of the field at
a given point. In this case as in ours, 
for each $C$ there
is a nontrivial classical solution. For a massive 
scalar field the solution would just 
be of the Yukawa potential form. The idea is to
perform the functional 
integral over fields first, in a saddle point approximation, before 
finally integrating over $C$. This procedure is well established 
in quantum mechanics and in the theory of instantons in gauge theories
with spontaneous symmetry breaking \cite{affleck}. 
In our case, we want ${\cal C}$ to be a 
local property of the geometry, which roughly speaking
measures how `spiky' it is. We define ${\cal C}$
on a surface $\Sigma_c$ of radius $a$ about a given point.
For fixed $a$ I shall show that 
such an instanton exists. Outside $\Sigma_c$ it
looks like one of 
Vilenkin's instantons. Inside it is just flat space.
As $a$ tends to zero, nothing in the definition 
becomes singular and  the Euclidean action remains 
finite
\footnote{
Strictly speaking, there is of course 
no need to take the 
limit as $a$ goes to zero. However, the description of the instantons
simplifies in that limit. This is particularly important in the
cosmological case \cite{ht1} where one is interested in
the Lorentzian continuation of the instanton. Only in the limit
as $a$ goes to zero will the continued spacetime be real.}.

Is
flat space is unstable against such instantons? I think this
is a nontrivial question. If the answer were positive,
one would I think be forced to discard the 
the path integral for quantum gravity as a sick theory.
Luckily, as I show, that turns out not to be the case.

How do we check for an instability? The signature is the 
imaginary contribution to the `self-energy' diagram for flat space,
mentioned above. The question is when we
integrate over these instantons and fluctuations about them,
whether any factors of $i$ emerge. If for fixed $C$ 
there is a negative mode in the field fluctuations, 
we  would be faced with
an integral of the form $\int dx e^{+x^2}$, which could
only be defined by rotating the contour. This would
introduce an $i$. The only other way we could get an $i$ is 
if the integral over $C$ performed at the end
of the calculation were not convergent, and required a similar
rotation to define it. For example, if we just inserted 
 $1 =\int dC \delta (C-x)$ into the above
integral over $x$, the $x$ integral would be no problem, but
the $i$ would reappear in the integral over $C$. 

For Vilenkin's instantons the 
second possibility is easy to exclude. There are no instantons
satisfying the boundary conditions at infinity for 
negative $C$.  But for positive $C$, the Euclidean 
action increases
linearly with $C$. So the integral over $C$ involves 
$\int_0^\infty dC e^{-C}$, which is perfectly convergent and
does not yield a factor of $i$. Physically, what this says is that
the Euclidean gravitational action, like the energy of
a rubber sheet, suppresses geometries with spikes in them.

The analysis for fluctuations about such instantons is quite technical.
As is well known, the 
Euclidean Einstein action is not positive semidefinite, so the 
path integral must be defined with care.
However, for
perturbations around instantons in pure gravity the procedure is 
well understood. One has to use a method which clearly
separates out the conformal mode
\cite{oldrefs},\cite{gpy},\cite{mottola}. 
I shall concentrate here on the simplest 
singular instanton described by Vilenkin, involving a massless
scalar field coupled to gravity. Such
instantons may
be reinterpreted {\it \`{a} la}  Kaluza-Klein as the Euclidean
Schwarzchild solutions of 
five dimensional gravity \cite{gar1}. As well as being
analytically known, the five metric
is regular, which allows a clearcut
analysis of fluctuations.
Furthermore,  because this is just pure gravity,
we can use well established gauge-fixing procedures
(\cite{gpy} and references therein).

I want to emphasise that I am {\it not} concerned here 
with the five dimensional
theory or its geometry.
I am studying four dimensional constrained instantons, and {\it only}
using the
five dimensional variables as a trick for 
gauge fixing in the path integral.
Nevertheless, from what is already  known about five dimensional 
gravity, there is a clear candidate for the negative mode 
and it only involves the variables 
of the four dimensional theory \cite{Hawking}. 
I shall construct it explicitly in order to prove that
it does not exist in the situation of interest. 

Witten argued that the five dimensional Euclidean Schwarzchild 
solution had a negative mode 
in analogy with the 
four dimensional case. The four dimensional version is
thought to describe the nucleation of 
black holes in hot flat space. In five dimensions Witten argued that
it
describes the decay of the Kaluza Klein vacuum. 
I shall not be concerned with any of these interpretations here,
nor with the five dimensional geometry. I merely 
use the five dimensional variables as a convenient choice for
performing the four dimensional functional integral.
The constrained variable
${\cal C}$ has a simple expression in terms of five dimensional 
variables,  and 
is not permitted to fluctuate in  the path integral over
fields. 
This condition turns out to eliminate 
the candidate negative mode mentioned above.
Without the
negative mode, the instantons do not yield an imaginary 
contribution to flat space to flat space amplitude. 
Flat space is therefore stable against decay via singular instantons.

\section{Vilenkin's Instantons}

Vilenkin's instantons 
are 
solutions to the field equations for 
four dimensional gravity coupled to a massless
scalar field.  They possess a singularity, 
and a conformal transformation reveals it to be
a boundary in the form of a three sphere of zero size\cite{ht1}. 
The boundary is perhaps more
disturbing than the presence of a singularity,
and  this 
might lead one to simply exclude such configurations 
from the path integral by fiat
\cite{Reall}. 
However, I shall now show that Vilenkin's instantons exist 
as a well defined limit 
of a 
regular class of constrained instantons,
with no boundary. So defined, they inevitably
contribute to the path integral.

\begin{figure}
\centerline{\psfig{file=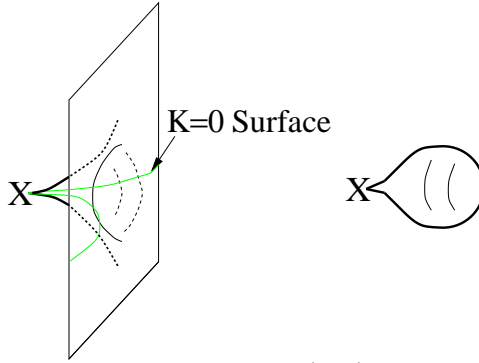,width=2.5in}}
\caption{Vilenkin's asymptotically flat instanton (flat)
is compared with the Hawking-Turok instanton (right).
Euclidean time runs upwards on Vilenkin's instanton.
If one identifies the lower
edge of the diagram (infinite negative Euclidean time) 
with flat three dimensional spacetime and cuts the
diagram horizontally through the singularity $X$ on  the three surface shown 
up on which $K_{ij}=0$,
and  continue Euclidean time to obtain a Lorentzian manifold with
a hole in it. Vilenkin claimed that if singular instantons were
permitted in the path integral, flat space would be unstable 
towards developing holes in this manner.
}
\labfig{vils}
\end{figure}

The action under consideration is usually written
\be
S_{E}= \int d^4x
\sqrt{g}\left[-{R\over  16 \pi G} + {1\over 2} (\partial \phi)^2 \right] -
\int_\Sigma  d^3x \sqrt{h} {K \over 8\pi G}.
\labeq{eact}
\ee
where last `surface' term is introduced to remove second derivatives
from the gravitational term. 
The induced three metric on 
$\Sigma$ is $h_{ij}$ and $K= K_{ij} h^{ij}$ is the trace of the second
fundamental form. In coordinates where $\Sigma$  is
a surface of constant $\tau$,
we have $K_{ij}= N^{-1}(-\partial_\tau h_{ij} +
N_{(i;j)})$ with $N$ and $N_i$ the lapse and shift functions. 
The surface term can be thought of as the rate of change of the proper
volume
of constant $\tau$ surfaces with respect to proper time.
That is, it is  the Euclidean version of
the Hubble constant times the three volume. 

We are interested in four manifolds with
no boundary other than that at infinity. The instantons of
interest will be
solutions of the field equations everywhere except on
a particular three surface $\Sigma_c$, where constraint is imposed. 
The action is a local integral of terms involving at most
first derivatives, and for the solutions of interest
the action density will be spread over space.
However, if we integrate by parts to write the action as
above, the bulk term actually vanishes since it is proportional to
the trace of the Einstein equations. But 
the constrained solutions will have discontinuous first derivatives
normal to $\Sigma_c$, and this leads to 
a contribution from the difference in
normal derivatives of the metric across $\Sigma_c$, as well
as a boundary term from the asymptotically flat surface at infinity.  

The constraint is defined as follows. We would like a variable which
measures the strength of `spikes' on the manifold. The natural 
one to 
choose is just
the surface term in the Einstein action. 
Around a  coordinate point on the manifold,  
draw a three surface $\Sigma_c$ 
of geodesic radius $a$. The constrained variable is then
\be
{\cal C} =
\[\int_{\Sigma_c}  d^3x \sqrt{h} {K \over 8\pi G}\]^+_-,
\labeq{constr}
\ee
where contributions from both sides of $\Sigma_c$ are included.
Each interpolating four geometry in the path integral will have some value for
${\cal C}$, and the 
functional delta function we introduce 
splits the path integral up accordingly. Only
geometries with discontinous first derivatives
possess nonzero ${\cal C}$. However this includes nearly all
geometries since 
the class of metrics with
continuous first derivatives is a set of measure zero.

Within each class of metrics so defined, there is a corresponding 
O(4) invariant instanton. If we write the metric in general
O(4) invariant form
\be
ds^2= n^2d\sigma^2 +b^2(\sigma)d \Omega_3^2
\labeq{emetric}
\ee
where $d \Omega_3^2$ is the metric for $S^3$,
then the action 
(\ref{eq:eact}) reduces to
\be
S_{E}= \int d \Omega_3 \int d\sigma \( -{3\over 8\pi G}(n^{-1}b b_\sigma^2 +nb) +
{1\over 2}  n^{-1} b^3 \phi_\sigma^2\).
\labeq{eacto}
\ee
The classical field equations are
\ba
(n^{-1} b^3 \phi_{\sigma})_\sigma = 0 \qquad 
b_\sigma^2 = {4\pi G\over 3}b^2 \phi_\sigma^2 +n^2.
\labeq{eqns}
\ea
where subscripts denote derivatives.
Our instantons will be solutions to the classical field
almost everywhere. The action is then given by
\be
S_{E}= - {2 \pi^2 \over 8 \pi G} n^{-1}(b^3)_\sigma|^\infty
+ \[{2 \pi^2 \over 8 \pi G} n^{-1}(b^3)_\sigma\]^+_-
\labeq{eactos}
\ee
where the second term is just the constrained variable (\ref{eq:constr}).

Now we discuss solutions to the field equations. We 
redefine 
$\sigma$ so that $n=1$. 
The scalar field equation possesses the general solution $ 
\phi_\sigma = A/b^3(\sigma)$, with $A$ an arbitrary constant. 
Asymptotic flatness requires that 
$b \sim \sigma$
at large $\sigma$. Thus $b$ satisfies 
\be
b_\sigma^2 - {4\pi G A^2 \over 3} b^{-4}=1,
\labeq{beq}
\ee
the equation for a particle of unit energy in 
a negative $b^{-4}$ potential. The exterior solution is unique
up to a constant shift in $\sigma$. 
Take our constraint surface to be
at $\sigma_c$.
The interior solution is flat space, 
with $b(\sigma)=(\sigma-\sigma_0)$ where $\sigma_0$ is the 
location of the origin in spherical coordinates, 
determined by matching $b$ at $\sigma_c$. By 
shifting $\sigma$ we can set  $\sigma_0=0$.
The scalar field is
constant in the interior region. 

With our instantons defined for finite $a$, we take the limit
as $a$ tends to zero. The exterior solution is that for the 
`singular instanton' solutions of \cite{ht1}, \cite{vil}.
The interior is flat space in Vilenkin's case, and nearly so
in the cosmological case. 
There is a contribution 
to the Einstein action from the difference (\ref{eq:eactos}) evaluated 
across $\sigma_c$. From (\ref{eq:beq}) in the limit of small $b$  
this contribution to the action is
$\sqrt{3\over 2} |A| 2\pi^2/\sqrt{8\pi G}$, which is strictly
positive, and increasing with $|A|$.
Additionally, there is a negative infinite contribution to the action 
from the surface at infinity. However in the path integral
one must normalise the one instanton contribution relative to
the no-instanton contribution. This means one has to subtract the
surface term appropriate to flat space, 
which is 
$-(2\pi^2/8 \pi G)\partial_R R^3 = - (2 \pi^2/ 8 \pi G)3  b^2$,
since at large distances we identify
$b$ with the radius $R$. The equation (\ref{eq:beq}) has
solution $b \sim \sigma + o(\sigma^{-3})$ at large $\sigma$, so
after subtraction the surface contribution from infinity is
actually zero.

To summarise: Vilenkin's instantons may be 
defined as a limit of
constrained instantons which are nonsingular (albeit with discontinuous
first derivatives)
and have no boundary. As such, they are legitimate contributions
to the path integral which must be present in quantum gravity
(and therefore presumably in the real world). But far from signalling
an instability, the Euclidean action monotonically increases as 
we increase the strength of the `spike'. 
As claimed in the introduction,
the fact that the action increases monotonically with $C$
guarantees that the
integration over $C$ is 
convergent, and no factors of $i$ emerge. This is just as for 
the surface of a balloon poked with a pencil: the energy increases.
If we take the pencil away the surface returns to being smooth.
In our case there is no pencil, but quantum mechanical 
vacuum fluctuations continually produce configurations
close to Vilenkin's instantons. They come and go, but 
never cause permanent damage.

\section{Fluctuations about Constrained Instantons}

As is well known, four dimensional gravity with a massless scalar
field can be obtained via dimensional reduction of
five dimensional gravity {\it \`{a} la} Kaluza Klein.
The five dimensional metric is given in terms of the four dimensional
metric and the scalar field as
\ba
g_{\mu \nu}^{(5)} = e^{\sqrt{2\over 3} {\phi\over M_{Pl}}} g_{\mu \nu}^{(4)}, \qquad
g_{5 5 }^{(5)} = e^{ -2 \sqrt{2\over 3} {\phi\over M_{Pl}}},\qquad g_{\mu 5}^{(5)}=0,
\labeq{ans}
\ea
where  $g_{\mu \nu}^{(4)}$  is the four 
metric in the Einstein frame, $\mu,\nu =0,1,2,3$ and 
$M_{Pl}^2 = (8 \pi G)^{-1}$.
The five dimensional field equations 
$R_{ab}^{(5)} =0$, $a,b=0...5$  reduce to those for a massless
field $\phi$ coupled to four dimensional gravity, and  the
$O(4)$ invariant 
solution described above is in fact
just the five dimensional 
Euclidean Schwarzchild solution. Unlike its Lorentzian 
counterpart, the 
latter is perfectly regular. It is intriguing that the
singularity of the four dimensional
metric $g_{\mu \nu}^{(4)}$ disappears when we 
change to the five dimensional one, but  
for present
purposes this is 
a mere calculational convenience.

We are interested in O(4) symmetric metrics, of the form
\ba
ds^2 = N^2(\tau)d\tau^2 +R^2(\tau) d\Omega_3^2 +r^2(\tau) d \phi^2.
\labeq{eso}
\ea
For the Euclidean Schwarzchild solution  
$N$, $R$ and $r$ are given by 
\ba
N_0&=&1 \qquad R_0^2 = C+\tau^2 \qquad r_0^2 = C \tau^2/(C+\tau^2).
\labeq{es}
\ea
but when we consider fluctuations we shall perturb $N$, $R$ and $r$.
$C$ is an integration constant, related to the `mass' of the
Schwarzchild solution ($R_0(\tau)$ is the usual Schwarzchild radial variable),
or in the Kaluza Klein interpretation
to the radius of the 
periodic  dimension at infinite $\tau$.

Comparing equations  (\ref{eq:emetric}), (\ref{eq:ans},) and 
(\ref{eq:eso}), we can construct the 5d to 4d dictionary,
\ba
r=e^{-\sqrt{2\over3} \phi/M_{Pl}} \qquad R= b e^{+{1\over 2}
 \sqrt{2\over3} \phi/M_{Pl}} \qquad 
N r^{1\over 2} d\tau = n d\sigma \qquad C^3= 16\pi G |A|^2/3
\labeq{dict}
\ea
The boundary term we are interested in is the four dimensional one, 
given in equation (\ref{eq:eactos}).
When expressed
in five dimensional variables, this is
\ba
\int d\Omega_3 {1 \over 8 \pi G_4}  
\[
N^{-1} r^{-{1\over 2}} \partial_\tau (R^3 r^{3\over 2} )\]_{\tau=0}. 
\labeq{boundary}
\ea
Note that this is {\it not} the boundary term for five dimensional
gravity - the latter would have the $r^{3\over 2}$ replaced by $r$
and $r^{-{1\over 2}}$ replaced by unity (cf. ref. \cite{gar1}).

\section{No Negative Mode}

We now consider perturbations of the background solution 
$g_{ab}^B$ discussed above. We set $g_{ab}=g_{ab}^B+h_{ab}$,
and compute the Euclidean 
Einstein action to second order in $h_{ab}$.
In addition to the problem of gauge fixing, the 
task of 
finding 
negative modes is  complicated by
the fact that the gravitational
action for conformal deformations of the
metric is unbounded below. At first sight there appear to 
be an infinite number of negative modes. However, these are unphysical.
In the context of pure gravity and for perturbations
around classical instantons the resolution of the problem has
been well understood for some time. One must be  
careful 
to separate the fluctuations of the conformal factor
from the 
transverse traceless fluctuations in the path integral
\cite{gpy} \cite{mottola}.

After a suitable gauge fixing term is added, the path integral over
the conformal factor decouples from that over the transverse traceless
metric perturbations. The latter are described by a quadratic
action involving the
Lichnerowicz operator, which involves the Riemann tensor
of the background solution. The 
problem then is to find the eigenmodes of the 
Lichnerowicz operator,
\ba
-\Box h_{ab} -2 R_{acbd}g_B^{ce} g_B^{df}h_{ef} = \lambda h_{ab}
\labeq{lichnero}
\ea
where the perturbation $h_{ab}$ is 
transverse and traceless, so $\Delta^a h_{ab}= g^{ab}_B h_{ab}=0$.
Here $\Delta^a$ and $\Box $ are
the usual covariant derivatives and Laplacian 
constructed from the background metric. 
If  equation (\ref{eq:lichnero}) has a normalisable
solution for negative $\lambda$, the instanton has a
genuine physical negative mode.

I shall consider only S-wave perturbations of the metric, since 
higher angular momentum perturbations are guaranteed to have greater
$\lambda$. The most general S-wave perturbation may be
written in terms of the variables $N$, $R$, and $r$ defined in 
(\ref{eq:eso}) and the background solution (\ref{eq:es}) as
\ba
N^2=1+2f \qquad R^2=R^2_0(1+2g) \qquad r^2=r^2_0(1+2h) 
\labeq{spert}
\ea
where tracelessness and transversality
read 
\ba
&& f+3g+h=0 \qquad 
\dot{f} +\gamma^{-1}\dot{\gamma}f
-3(\dot{R_0}/R_0)g -(\dot{r_0}/r_0)h =0.
\labeq{tt}
\ea
and (\ref{eq:lichnero}) reads
\ba
-\ddot{f}-\gamma^{-1}\dot{\gamma}f +6\[ {\ddot{R_0} \over R_0}
-\({\dot{R_0}\over R_0}\)^2 \]g +2 \[ {\ddot{r_0} \over r_0}
-\({\dot{r_0}\over r_0}\)^2 \]h +\[6\({\dot{R_0}\over R_0}\)^2 +
2 \({\dot{r_0}\over r_0}\)^2 \] f = \lambda f
\labeq{perteq}
\ea
where $\gamma = R_0^3 r_0$.
Equations (\ref{eq:tt}) 
can be used to
eliminate $g$ and $h$ from (\ref{eq:perteq}). Using  
$R_0(\tau)$ and $r_0(\tau)$ from (\ref{eq:es}) we find
\ba
-\ddot{f}-{3+10 \tau^2 -5\tau^4 \over \tau (1-\tau^4)}\dot{f}
-{20 \over (1-\tau^4)} f =\lambda f.
\labeq{perteqa}
\ea
This equation possesses regular singular points at $\tau=0$,
$1$ and $\infty$. Regularity at $\tau=1$ requires that
$\dot{f}(1)=-{5\over 2} f(1)$. One can use a shooting technique
to search for the lowest eigenvalue $\lambda$. 
One starts from $\tau=1$ with $f=1$ and $\dot{f}=-{5\over 2}$.
Given $\lambda$,  the solution is 
propagated to large $\tau$ by solving the differential equation.
One adjusts $\lambda$ so that the
solution goes to zero at infinite $\tau$. Having found $\lambda$,
the solution for $\tau <1$ is found by solving the 
equation with boundary condition $\dot{f}(0)=0$.
The damping term in the equation is singular and 
ensures that
as $\tau$ approaches unity the correct relation between $\dot{f}$ and
$f$ is satisfied. Finally, rescaling the $0<\tau<1$  part of
the solution to match the $\tau>1$ part at $\tau=1$, 
one has the complete eigenfunction. With this procedure 
I found only one negative value for $\lambda$, namely $\lambda
= - 1.25$, with small error. 
The negative mode is shown in Figure 6.

\begin{figure}
\centerline{\psfig{file=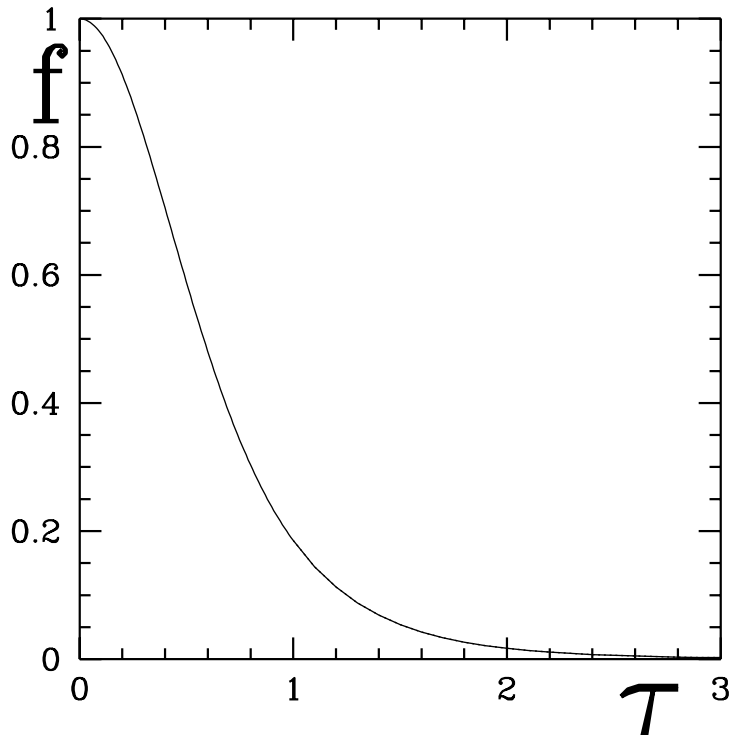,width=4.in}}
\caption{The negative mode for the five dimensional Euclidean
Schwarzchild solution. This mode is eliminated from
the four dimensional theory by the
constraint required for the existence of the
instanton.}
\labfig{bound}
\end{figure}

Now we have discovered the single allowed negative mode, the
key question is whether it is allowed by the constraint
on (\ref{eq:boundary}). 
When we perturb the instanton, only  perturbations
leaving (\ref{eq:boundary}) unperturbed are allowed.
This condition reads
\be
\delta [N^{-1} r^{-{1\over 2}} \partial_\tau (R^3 r^{3\over 2} )]_{\tau=0}
\propto 
[3 g +h-f]_{\tau=0} =0,
\labeq{boundaryp}
\ee
where I used 
$r_0 \sim \tau$ and  $R_0 \sim$ const as
$\tau$ goes to zero.
However the traceless condition imposes $f+3g+h=0$ 
and transversality imposes $f=h$ at $\tau=0$ (see (\ref{eq:tt})).
These conditions together require $f=0$ at $\tau=0$. 
The negative mode we have found, which is the only one, is
therefore excluded.
Note that the 
constraint on (\ref{eq:boundary}) 
has no effect on any of the higher
$S^3$-dependent modes, since these give no contribution
to the boundary term 
when integrated over $\Sigma$.

I conclude that if the instantons under consideration 
are properly treated as constrained instantons, as
they must be,
they do not possess a negative mode and therefore
do not lead to the decay of flat spacetime.
I believe the calculations reported above should be interpreted
as a nontrivial test of the quantum gravitational
path integral, and as such they are a good sign. Furthermore, the 
above discussion of constrained instantons does, I think, 
considerably clarify their interpretation in the cosmological
context \cite{ntprep}.

\medskip
\centerline{\bf Acknowledgements}

I am indebted to M. Bucher, 
S. Gratton, H. Reall, S.W. Hawking, M. Perry 
and T. Wiseman  for helpful 
discussions of this problem.


\end{document}